\documentclass[letterpaper]{JHEP3}
\usepackage{graphicx}
\usepackage{epsfig}
\newcommand{\tg}{\ensuremath T_\gamma}
\newcommand{\tr}{\ensuremath T_R}
\newcommand{\mev}{\ensuremath \mathrm{~MeV}}
\newcommand{\kev}{\ensuremath \mathrm{~keV}}
\newcommand{\sstt}{\ensuremath \sin^22\theta}
\newcommand{\onod}{\ensuremath \Omega_{\nu s}/\Omega_\mathrm{dm}}
\newcommand{\gp}{\ensuremath \Gamma_\phi}
\title{Sterile neutrino production in models with low reheating temperatures}
\author{Carlos E. Yaguna\\ Department of Physics and Astronomy, UCLA, 475
  Portola Plaza, Los Angeles, CA 90095,
  USA\\E-mail:\email{ yaguna@physics.ucla.edu}}
\abstract{By numerically solving the appropriate Boltzmann
equations, we study the  production of sterile neutrinos
in models with low reheating temperatures. We take into account the production in
oscillations as well as in direct decays  and compute the sterile neutrino primordial spectrum, the effective
number of neutrino species, and the sterile neutrino contribution to the
mass density of the Universe as a function of the mixing and the reheating
parameters. It is shown that sterile neutrinos with non-negligible mixing
angles do not necessarily  lead to  $N_\nu\sim 4$ and that sterile neutrinos may have the right relic density to explain the
dark matter of the Universe. If dark matter consists of sterile neutrinos
produced in oscillations, X-rays measurements set a strong limit on the
reheating temperature, $\tr\gtrsim 7\mev$. We also point out that the direct decay
opens up a new production mechanism for  sterile
neutrino dark matter where cosmological constraints can be satisfied.}
\preprint{}
\begin{document}
\section{Introduction}
Sterile neutrinos likely exist. They can easily be incorporated into the
standard model and provide the simplest explanation for the existence of neutrino masses. The most
important parameter associated with sterile neutrinos is probably their mass
scale. In seesaw models \cite{seesaw}, where sterile neutrinos are simply added to the
standard model matter fields in order to generate light neutrino masses,
sterile neutrino masses are free parameters of the Lagrangian, whose values
are to be experimentally determined. To account for the neutrino masses inferred
from  the solar and atmospheric neutrino experiments,  at least two
sterile neutrinos are required but only mild constraints on the masses or
mixing of the sterile neutrinos can be derived. And theoretical considerations
are of no help either, for  heavy as well as light sterile
neutrinos can be  motivated on different grounds \cite{deGouvea:2005er,kusenko}. It seems reasonable, then,
to consider the sterile neutrino mass scale simply as  another free parameter
subject to present experimental constraints. In this paper, we
study sterile neutrinos with masses in the $\mathrm{eV}$--$\mathrm{keV}$ range.

Sterile neutrinos with keV masses have indeed been proposed as dark matter
candidates \cite{dm1,dm2,dm3}. In the early Universe, such sterile
neutrinos are produced in active-sterile neutrino oscillations and never reach
thermal equilibrium. Due to their  primordial velocity
distribution, sterile neutrinos damp inhomogeneities on small scales and
therefore behave as warm dark matter particles. The mass of dark matter sterile
neutrinos is constrained from below by the  observed clustering on small scales
of the Lyman-$\alpha$ forest \cite{Hansen:2001zv}. Present  bounds give
$m_s> 10\mbox{-}14 \kev$ \cite{Seljak:2006qw,Viel:2006kd}. Because of its mixing with active neutrinos, the
$\nu_s$ may radiatively  decay (through $\nu_s\to \nu + \gamma$) producing a monoenergetic photon
with  $E_\gamma\sim m_s/2$. X-rays measurements may therefore be used to constraint or
infer the mass of the sterile neutrino.  Recent bounds, based on
observations of the Virgo and Coma clusters and the X-ray background, yield $m_s<6\mbox{-}10\kev$ \cite{Abazajian:2006yn,Boyarsky:2005us,Boyarsky:2006zi} and are thus in
conflict with the Lyman-$\alpha$ forest constraint. That means that the
minimal  mechanism for sterile neutrino dark matter, based on active-sterile
oscillations, is already
ruled out \cite{Abazajian:2006yn,Seljak:2006qw,Viel:2006kd}.

A possible clue regarding the mass scale of the sterile neutrinos
is the result of the LSND experiment \cite{Athanassopoulos:1996jb}. It found evidence
of $\bar{\nu_\mu}\to\bar{\nu_e}$ conversion, which is  being tested
by the Fermilab MiniBoone experiment \cite{Stancu:2006gv}. The LSND signal can be explained by the
existence of light ($m_s\sim 1-10$~eV) sterile neutrinos mixed with $\nu_e$
and $\nu_\mu$ \cite{deGouvea:2005er}. In the standard cosmological model, such
sterile neutrinos generate two important problems: i) They give a
contribution to $\Omega_\nu$ larger than that suggested by global fits of CMD
and LSS data \cite{Pierce:2003uh}. ii) They thermalize in the early Universe
so that $N_\nu\sim 4$, in possible conflict with big-bang
nucleosynthesis bounds \cite{Abazajian:2002bj}. Recently, the MiniBoone
experiment presented its first results \cite{Aguilar-Arevalo:2007it}
 which disfavore even more the
so-called (3+1) schemes \cite{Maltoni:2007zf}. It seems, nonetheless, that
(3+2) schemes are still viable \cite{Maltoni:2007zf}.

The standard cosmological model, however,  has not been tested beyond big
bang nucleosynthesis, for $T\gtrsim 1 \mev$. Cosmological models  with low reheating
temperatures, for example, offer a natural and viable alternative to the
standard paradigm. In fact, various scenarios of physics
beyond the standard model, including supersymmetry and superstring theories,
predict the existence of massive particles with long lifetimes that decay
about the big bang nucleosynthesis epoch, inducing a low reheating
temperature and modifying the initial conditions of
the standard cosmology. Over the years, different issues related to these
models have been studied in the literature
\cite{Kawasaki:2000en,Giudice:2000ex,Hannestad:2004px}. In this paper we
consider the possible interplay between sterile neutrinos and models with low
reheating temperatures. On the one hand, sterile neutrinos may serve as probes of
the early Universe and constrain the reheating temperature. On the other hand, models with low reheating temperatures may
alleviate some of the problems associated with sterile neutrinos, suppressing their abundance  or modifying the standard
relation between the sterile neutrino  relic density and the mixing parameters.

So far, a detailed analysis of these effects have not been presented in the
literature. Cosmologies with low reheating temperatures were suggested, in
\cite{Abazajian:2002bj}, as a possible way to accommodate the LSND signal and
big bang
nucleosynthesis, whereas in \cite{Gelmini:2004ah}, several simplifying assumptions -not all
of them justified- were used to obtain and  analytic estimation of the sterile
neutrinos produced in oscillations. In this paper, we numerically solve the
equations that determine the sterile neutrino distribution function in models
with low reheating temperatures. Two different sources of
sterile neutrinos are taken into account: active-sterile oscillations and  the direct decay of the field
responsible for the reheating process. We compute different observables
related to  the sterile neutrino,
including its spectrum and  relic density, as a function of the
reheating parameters and the mixing angle and mass of the sterile neutrino.  

In the next section we describe the reheating process and introduce the different equations that are relevant for the production of
sterile neutrinos. Then, the  behavior of active neutrinos in models with low reheating
temperatures will be briefly reviewed. In section \ref{sec:4}, we study in detail
the production of sterile neutrinos as a result of active-sterile neutrino
oscillations for different mixing and reheating parameters. We show that
$N_\nu\sim 3$ can be obtained even
for sterile neutrinos with relatively large mixing angles and that dark matter
sterile neutrinos provide a strong constraint on the reheating temperature. Finally, in
section \ref{sec:5}, we include  the production of sterile neutrinos through
the direct decay of the scalar field and study the resulting sterile neutrino
spectrum and relic density. We observe that sterile neutrinos produced in
decays may account for the dark matter and avoid the Lyman-$\alpha$ and X-ray constraints.

\section{The reheating process}
Reheating is defined as the transition period between a Universe dominated by
a unstable non-relativistic particle, $\phi$, and the radiation dominated Universe. In
the standard cosmological model reheating is assumed to occur only after
inflation, but in general, additional reheating phases not related to
inflation are also possible and our discussion applies equally to them. During reheating the dynamics of the Universe is rather involved. The energy
density per comoving volume of the non-relativistic particle decreases as
$e^{-\Gamma_\phi t}$ -with $\Gamma_\phi$ the $\phi$ decay width- whereas the
light decay products of the $\phi$ field thermalize. Their temperature
quickly reaches a maximum value $T_{max}$ and then decreases as $T\propto
a^{-3/8}$ \cite{Giudice:2000ex}, as  a result of the continuous entropy release. During this time the relation between the expansion
rate and the temperature is neither that of a matter-dominated universe
($H\propto T^{3/2}$) nor that of a radiation-dominated Universe ($H\propto
T^4$) but it is given instead by $H\propto T^4$. Thus, at a given temperature
the Universe expands faster during reheating than in the radiation-dominated
era. This unusual behavior continues until $t\sim \Gamma_\phi^{-1}$, when the
radiation dominated phase commences with temperature $\tr$. From then on, that is for $T<\tr$, the evolution of the Universe proceeds as in the
standard scenario but with initial conditions determined by the reheating
process.

The success of  standard big bang nucleosynthesis provides the strongest
constraint on $\tr$. Electrons and photons interact electromagnetically
and consequently have large creation,
annihilation and scattering rates that  keep them in equilibrium
even during reheating. Neutrinos, on the contrary, can interact only through
the weak interactions and are slowly produced in electron-positron
annihilations.  Since big bang nucleosynthesis requires a thermal neutrino
spectrum, $\tr$ should be high enough to allow the thermalization of the neutrino sea.  Given that,  in the standard
cosmology, neutrinos decouple from the thermal plasma at $T\sim 2-3 \mev$, it
can be estimated that they will not thermalize if $\tr<$ few$\mev$. Indeed,
detailed calculations  give $T\gtrsim 2-4 \mev$
\cite{Kawasaki:2000en,Hannestad:2004px} as the present bound. In this paper, we consider
models with reheating temperatures below $10\mev$.

Let us know formulate the equations that describe the reheating process, and
in particular, the production of sterile neutrinos at low reheating
temperatures. We denote by $\phi$ the unstable non-relativistic particle that
initially dominates the energy density of the Universe. Its energy
density, $\rho_\phi$, evolves according to 
\begin{equation}
\frac{d\rho_\phi}{dt}=-\gp\rho_\phi- 3H\rho_\phi
\end{equation}
where $H$ is the Hubble parameter and $\gp$ is the $\phi$ decay width.

The energy-momentum conservation equation in the expanding universe is 
\begin{equation}
\frac{d\rho_T}{dt}=-3 H(\rho_T+P_T)
\label{eq:cons}
\end{equation}
where $\rho_T$ and $P_T$ denote respectively the total energy density and the
total pressure. At the low  temperatures we allow for, only the scalar field,
electrons, photons, and neutrinos are present in the plasma. Denoting by
$\rho_\nu$ the energy density in active and sterile neutrinos, we have that
\begin{equation}
\rho_T(t)=\rho_\phi+\rho_\gamma+\rho_e+\rho_\nu
\end{equation}
 and an analogous expression holds for $P_T$. Equation (\ref{eq:cons}) can be
 rewritten as an evolution equation for the (photon) temperature as
\begin{equation}
\frac{d\tg}{dt}=-\frac{-\rho_\phi\gp+4H\rho_\gamma+3H(\rho_e+P_e)+4H\rho_\nu+d\rho_\nu/dt}{\partial\rho_\gamma/\partial\tg+\partial\rho_e/\partial\tg}\,.\end{equation} 
$H$, the hubble parameter, is given by the Friedmann equation,
\begin{equation}
H(t)=\frac{\dot{a(t)}}{a(t)}=\sqrt{\frac{8\pi}{3}\frac{\rho_T}{M_P^2}}
\end{equation}
with $a$ the scale factor and $M_P$ the Planck mass .

We follow the evolution of active neutrinos by solving the momentum-dependent
Boltzmann equation
\begin{equation}
\frac{\partial f_\nu}{\partial t}- Hp\frac{\partial f_\nu}{\partial p}=C_{coll}
\label{eq:bolt}
\end{equation}
for $\nu_e$ and $\nu_\mu$ ($f_{\nu_\tau}=f_{\nu_\mu}$). $C_{coll}$, the total
collision term,  describes
 neutrino annihilations and scatterings. The following processes are taken
 into account in our calculations:
\begin{eqnarray}
\nu_i+\nu_i&\leftrightarrow& e^++e^-\\
\nu_i+e^\pm&\leftrightarrow& \nu_i+e^\pm\,.
\end{eqnarray}
The collision terms associated with these processes are complicated, involving
nine-dimensional integrations over momentum space. But they can be
simplified to one-dimensional integrals  by  neglecting $m_e$ and assuming that electrons
obey the Boltzmann distribution \cite{Kawasaki:2000en}. Since the error due to the above
approximations is small (less than few percent), we will use the
one-dimensional form of the collision terms.

Regarding the sterile neutrinos, we will  consider the simplifying limit of
two neutrino (active-sterile) mixing. That is, we assume one sterile neutrino,
$\nu_s$, that mixes predominantly with a single active flavor $\nu_\alpha$
($\alpha=e,\mu,\tau$). In
consequence, the transformation between the flavor and the mass bases can be
written as
\begin{eqnarray}\label{eq:mix}
|\nu_\alpha\rangle&=&\cos\theta\,|\nu_1\rangle+\sin\theta\,|\nu_2\rangle\\
|\nu_s\rangle&=&-\sin\theta\,|\nu_1\rangle+\cos\theta\,|\nu_2\rangle
\end{eqnarray}
where $|\nu_1\rangle$ and $|\nu_2\rangle$ are neutrino mass eigenstates with
masses $m_1$ and $m_2$, respectively. $\theta$, the mixing angle, parameterizes
the magnitude of the mixing between the active and the sterile neutrino. For
the small mixing angles we deal with, $|\nu_2\rangle$ practically coincides
with $|\nu_s\rangle$, so we will use $m_s$ instead of $m_2$ to denote the mass
of the eigenstate that is predominantly sterile.  

The sterile neutrino distribution function also follows a Boltzmann equation
like (\ref{eq:bolt}). The collision term for
$\nu_\alpha\leftrightarrow \nu_s$ oscillations is \cite{dm2}:
\begin{equation}
C_{\nu_s\leftrightarrow \nu_\alpha}=\frac{1}{4}\frac{\Gamma_\alpha(p)\Delta^2(p)\sstt}{\Delta^2(p)\sstt+D^2(p)+\left[\Delta(p)\cos2\theta-V^T(p)\right]^2}\left[f_\alpha(p,t)-f_s(p,t)\right]
\label{eq:nbol}
\end{equation}
where $\Delta(p)=m_s^2/2p$, $\Gamma_\alpha$ is the $\nu_\alpha$ total
interaction rate, $D(p)=\Gamma_\alpha/2$ is the quantum damping rate, and
$V^T$ is the thermal potential.

In addition to oscillations, we also consider the production of sterile
neutrinos through the direct decay $\phi\to \nu_s\nu_s$. Since $\phi$ is
nonrelativistic, each sterile neutrino is born with momentum $m_\phi/2$ and
the collision integral becomes
\begin{equation}
C_{\phi\to \nu_s\nu_s}= b\frac{2 \pi^2}{(m_\phi/2)^2}\Gamma_\phi n_\phi
\delta(p-m_\phi/2)\,,
\label{eq:dec}
\end{equation}
where $b$ is the branching ratio into sterile neutrinos,  and
$m_\phi\,,n_\phi$ are respectively  the $\phi$ mass and number density.

As initial conditions we assume that at early times  the energy-density of the
Universe is dominated by $\phi$, and that  active
and sterile neutrinos are  absent from the primordial plasma. As long as the maximum temperature reached by the
plasma ($T_{max}$ \cite{Giudice:2000ex}) is large enough, the final outcome is independent of the
initial conditions. We found that $T_{max}\sim 20 \mev$ is enough to guarantee
such independence.

Our analysis can naturally be divided into two parts: production in
oscillations only ($b=0$), and production in oscillations and decay ($b\neq
0$). In the first case, to be investigated in section \ref{sec:4}, the parameters that enter into the above
equations are $m_s$, $\sstt$, and  $\Gamma_\phi$. It is customary to trade
$\Gamma_\phi$ with the cosmological parameter $\tr$ -known as the reheating temperature- through the relations
\begin{equation}
\Gamma_\phi= 3 H(\tr)
\label{eq:h1}
\end{equation}
and 
\begin{equation}
H(\tr)=3\frac{\tr^2}{M_P}\left(\frac{8 \pi^3g_*}{90}\right)^{1/2}\,.
\label{eq:h2}
\end{equation}
with $g_*=10.75$. These equations establish a one-to-one correspondence
between $\Gamma_\phi$ and $\tr$. In the second case, when sterile neutrinos are also produced in
decays ($b\neq 0$), the results will depend additionally  on $b$ and
$m_\phi$. Section \ref{sec:5} deals with this interesting possibility.

For a given set of mixing and reheating parameters, we simultaneously follow
the evolution of $\rho_\phi$, $\tg$, $f_{\nu_e}(p)$, $f_{\nu_\mu}(p)$, and
$f_{\nu_s}(p)$ from the matter dominated era well into the radiation-dominated
Universe, until the distribution functions reach their
asymptotic values (\mbox{$T<0.1\mev$}).  The main output from this system of equations are the
neutrino distribution functions, which can be used to compute several
observables. Big bang nucleosynthesis, for instance, is sensitive to the
relativistic energy density in neutrinos. This quantity is usually   parameterized
in units of the energy density of a standard model neutrino, $\rho_{\nu_0}$,
and denoted by $N_\nu$,
\begin{equation}
N_\nu=\frac{\rho_{\nu_e}+\rho_{\nu_\mu}+\rho_{\nu_\tau}+\rho_{\nu_s}}{\rho_{\nu_0}}\,.
\end{equation}
Since sterile neutrinos are dark matter candidates, it is also important to
compute their relic abundance,
\begin{equation}
\Omega_s=\frac{m_sn_s}{\rho_c}\,,
\end{equation}
where $m_s\,,n_s$ are respectively the mass and number density of the sterile
neutrinos, and $\rho_c$ is the critical density of the Universe.

\section{Active neutrinos and low $\tr$}

The  evolution of the sterile neutrino distribution function strongly depends on the
corresponding function of the active neutrino flavor with which it mixes and it is in many ways analogous to
it. Before considering sterile neutrinos, it is therefore appropriate to briefly review the salient features
related to the behavior of active
neutrinos in models with low reheating temperatures. 

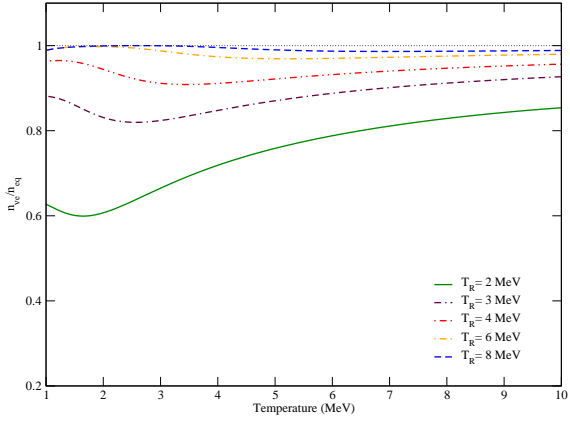
\DOUBLEFIGURE[t]{evolacte,width=7.5cm}{evolactm,width=7.5cm}{The evolution of the electron
  neutrino number density as a function of the photon temperature for different
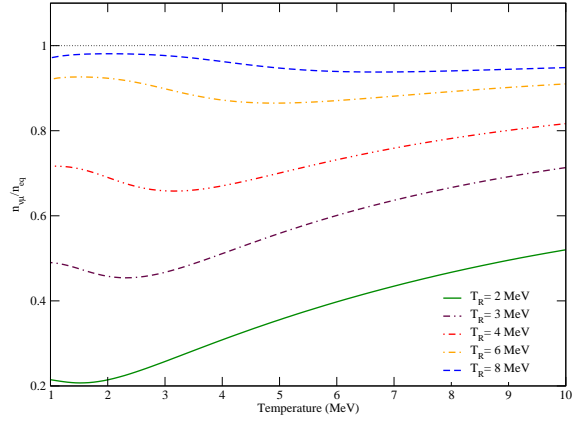
  reheating temperatures.\label{fig:1}}{The evolution of the muon (or tau)
  neutrino number density as a function of the photon temperature for different
  reheating temperatures.\label{fig:2}}

Figure \ref{fig:1} shows the evolution of the electron neutrino number
density (normalized to the equilibrium density) as a function of the
temperature for different reheating temperatures. The pattern is
clear. At high temperatures, $T\gg \tr$, neutrinos are  out of equilibrium and
$n_{\nu_e}/n_{eq}$ continually decreases with time until $T\sim \tr$ is
reached. For $T<\tr$, neutrinos evolve as in the radiation dominated but with
a non-equilibrium initial condition ($n_{\nu_e}(\tr)\neq n_{eq}(\tr)$). If
$\tr$ is large enough, neutrinos will be able to recover the equilibrium
distribution before  decoupling from the thermal plasma. Such event, illustrated by the line
$\tr=8 \mev$ in figure \ref{fig:1},  would be
indistinguishable from the standard cosmology. For smaller reheating
temperatures, on the other hand, neutrinos never reach the equilibrium
distribution and decouple from the plasma with a smaller abundance than in the standard scenario. That is
exactly what happens, for instance, if $\tr\lesssim 4\mev$ (see figure
\ref{fig:1}). Note nonetheless that even for $\tr=3\mev$ the asymptotic deviation from the
standard prediction amounts to less than $10\%$.

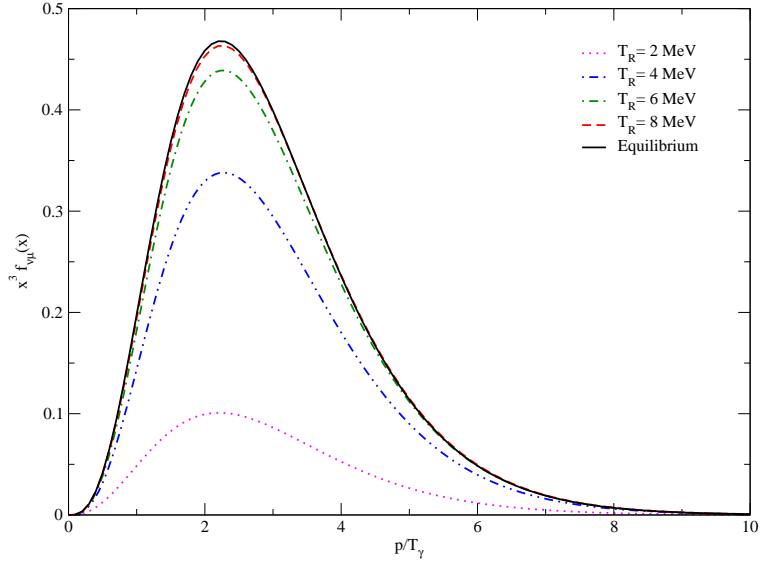
\EPSFIGURE[b]{specactm,width=10cm}{The primordial energy spectrum of the muon neutrino
  as a function of $p/\tg$ for different reheating temperatures.\label{fig:3}}

Because muons  are not present in the thermal plasma at low temperatures, muon
neutrinos can only be produced in neutral-current interactions. Consequently,
the muon neutrino  deviates from equilibrium farther than the
electron neutrino, as
revealed in figure \ref{fig:2}. Indeed, for $\tr=3\mev$ the deviation from
the standard prediction amounts to $50\%$.

The effects of the reheating process can also be seen in the primordial neutrino
spectrum. A equilibrium spectrum with
$T_\nu=\tg/1.4$ is expected in the standard cosmological model.  Figure
\ref{fig:3} shows the $\nu_\mu$ primordial
energy spectrum for different values of $\tr$  as a function of $p/\tg$. The deviation from equilibrium is
clearly visible for the smaller reheating temperatures.

\section{Sterile neutrino production in oscillations}\label{sec:4}
Let us now consider the production of sterile neutrinos through active-sterile
neutrino oscillations. For simplicity we will consider mixing with the
electron neutrino only so that  $\sstt$ denotes
the mixing angle between $\nu_e$ and $\nu_s$. We are then left with $3$
parameters that determine all the observables: $\tr$, $\sstt$, and $m_s$. In
this section we study how these parameters affect $f_{\nu_s}$, $N_\nu$, and
$\Omega_{\nu_s}$. 

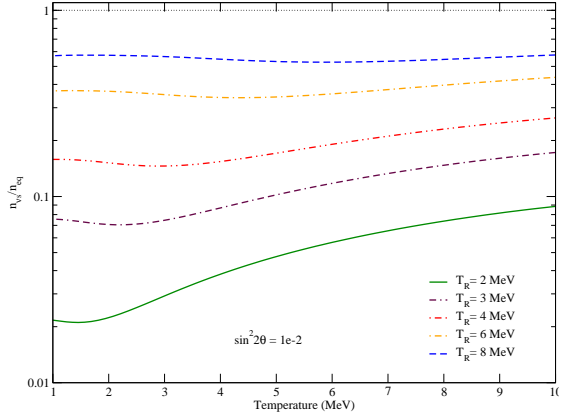
\DOUBLEFIGURE[t]{evolsttr,width=7.4cm}{evolsttr4,width=7.4cm}{The evolution of the
  sterile neutrino number density as a function of the photon temperature for different
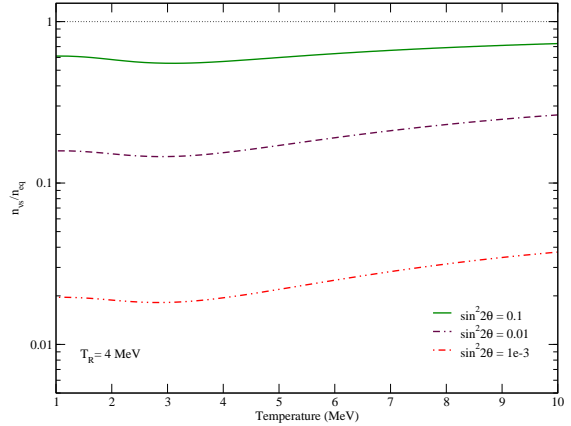
  reheating temperatures and  $\sstt=10^{-2}$.\label{fig:5}}{The evolution of the
  sterile neutrino number density as a function of the photon temperature for different
  mixing angles and $\tr=4 \mev$.\label{fig:4}}

The evolution of the sterile neutrino number density follows a pattern
similar to that of the active neutrinos.
Figure \ref{fig:4} shows  $n_{\nu_s}/n_{eq}$
as a function of the temperature for different values of $\tr$ and
$\sstt=10^{-2}$. Sterile neutrinos are always out of equilibrium and 
$n_{\nu_s}/n_{eq}$ decreases with time during the reheating phase, reaching its minimum value at
$T\sim \tr$. At $T\lesssim \tr$,  the universe is radiation dominated and the
sterile neutrino
population slightly increases, in part as a result of the corresponding increase in
$n_{\nu_e}$ (see figure \ref{fig:1}). The
asymptotic value of $n_{\nu_s}/n_{eq}$, however, differs very little from its
value at $\tr$.

Note that this result is  at odds with \cite{Gelmini:2004ah}, where it was \emph{assumed} that the
production of sterile neutrinos starts at $\tr$. Actually, as we have seen, sterile neutrinos are slowly
created during the $\phi$ dominated era and only a small fraction of them are
produced after $\tr$.

For the range of sterile neutrino masses considered,  $n_{\nu_s}/n_{eq}$ does
not depend on $m_s$. Thus, the other relevant dependence to investigate
is that with $\sstt$. In figure \ref{fig:5}, $n_{\nu_s}/n_{eq}$ is shown
as a function of the temperature for $\tr=4\mev$ and different mixing
angles.  As expected,  the smaller the mixing angle the smaller
$n_{\nu_s}/n_{eq}$. Indeed, for small mixing angles ($\sstt\lesssim 10^{-2}$),
$n_{\nu_s}/n_{eq}\propto \sstt$, as seen in figure \ref{fig:5}. Such proportionality
is expected when $f_{\nu_s}$ can be neglected with respect to $f_{\nu_e}$ in
equation (\ref{eq:nbol}). At large
mixing angles  $f_{\nu_s}$ may  become comparable with
$f_{\nu_e}$ and the above relation no longer holds. Neglecting $f_{\nu_s}$ in
(\ref{eq:nbol}), therefore, is not a good approximation for  sterile neutrinos
with large mixing angles.

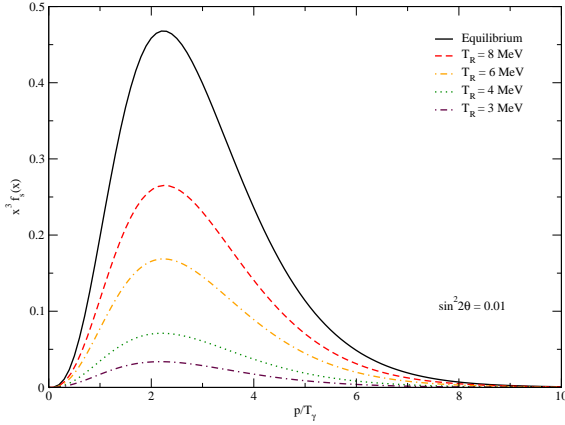
\DOUBLEFIGURE[t]{specsttr,width=7.5cm}{specsttr4,width=7.5cm}{The primordial energy spectrum of the sterile neutrino
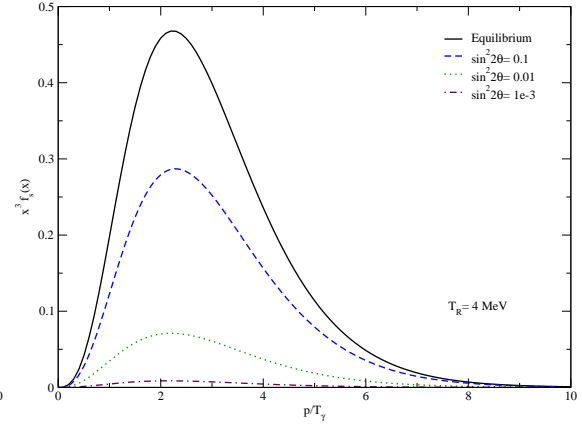
  as a function of $p/\tg$ for different reheating temperatures  and $\sstt=10^{-2}$.\label{fig:6}}{The primordial energy spectrum of the sterile neutrino
  as a function of $p/\tg$ for different mixing angles and $\tr= 4\mev$.\label{fig:7}}

The primordial energy spectrum of the sterile neutrino is shown in
figures \ref{fig:6} and \ref{fig:7} for  different values of $\tr$  and $\sstt$. It is
certainly non-thermal and is strongly suppressed for low reheating
temperatures or small mixing angles.

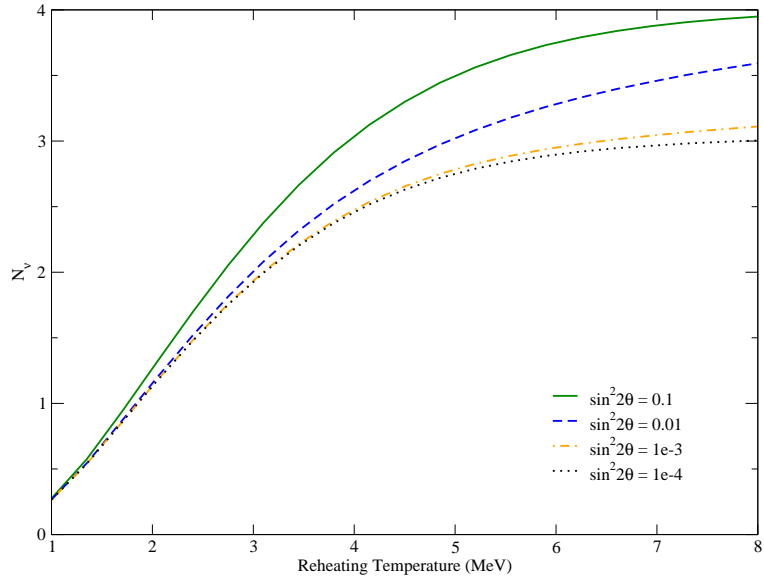
\EPSFIGURE[b]{nntrhe,width=10cm}{The effective number of neutrino species as a
  function of $\tr$ for different mixing angles.\label{fig:8}}

Standard big bang nucleosynthesis is  a powerful cosmological
probe of active and sterile neutrino effects. It constrains the number of
thermalized neutrinos present at $T\sim 0.1-1\mev$ to be $N_\nu=2.5\pm
0.7$ \cite{Cirelli:2004cz}. Unfortunately, the uncertainty in $N_\nu$ is
controversial so not strict bound on it  can be derived. Here, we will simply take
as a reference value the prediction of the standard
cosmological model, $N_\nu=3$. Figure \ref{fig:8} shows  $N_\nu$ as a function of $\tr$ for different mixing angles. The variation with
$\tr$ is strong, going from $N_\nu\sim 3-4$ for $\tr\gtrsim 7\mev$ to $N_\nu\sim
0.3$ for $\tr=1\mev$. The spread due to different mixing angles, on the other
hand, is maximum ($\Delta N_\nu\sim 1$) at large $\tr$, and  decreases for
smaller $\tr$. Note that for $\sstt\lesssim 10^{-3}$, $N_\nu$ is essentially
insensitive to the presence of sterile neutrinos; it becomes a function only of
$\tr$.  As expected, the standard cosmological scenario is recovered at
large $\tr$.  In that region, if the mixing angle is large $\sstt\sim 0.1$ all
neutrinos -the three active plus the sterile- thermalize, yielding $N_\nu\sim
4$. That is not necessarily the case for lower reheating temperatures,
however. If $\tr\sim 4\mev$, for instance, then $N_\nu\sim 3$ for a sterile
neutrino with $\sstt\sim 0.1$; and the same $N_\nu$ can be obtained for
$\sstt\sim 10^{-2}$ and $\tr=5\mev$. Hence,  LSND sterile neutrinos  may still yield $N_\nu\sim 3$, avoiding possible
conflicts with big bang nucleosynthesis.

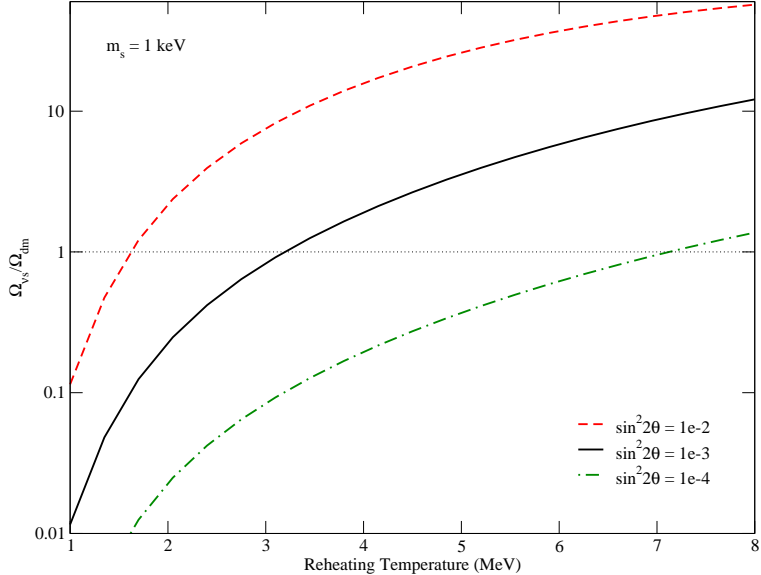
\EPSFIGURE[t]{omegatrh,width=10cm}{$\onod$ as a function of $\tr$ for different
  mixing angles and $m_s= 1\kev $.\label{fig:9}}

The sterile neutrino relic density as a function of $\tr$ is shown in figure
\ref{fig:9} for different mixing angles and $m_s=1\kev$. Along the horizontal
line, sterile neutrinos entirely account for the dark matter density of the Universe. The
region above the horizontal line is therefore ruled out, whereas below it, $\nu_s$ only
partially contribute to the dark matter density. Thus, in the region
$3\mev<\tr <7 \mev$ and $10^{-3}> \sstt >10^{-4}$ a sterile
neutrino with $m_s=1\kev$ may explain the dark matter.

Because $\Omega_{\nu_s}$ scales linearly with $m_s$, the results for a
different value of $m_s$ can easily be obtained from the same figure. First  notice 
from the figure that the sterile neutrino relic density also depends linearly on $
\sstt$. So, another region where $\Omega_{\nu_s}=\Omega_{dm}$ is
$m_s=10\kev$, $3\mev<\tr <7\mev$ and $10^{-4}>\sstt >10^{-5}$. 

\EPSFIGURE[t]{omesin,width=10cm}{The sterile neutrino relic density as a
  function of $\sstt$.\label{fig:10b}}

In the standard cosmological scenario, dark matter sterile neutrinos are
produced at $T\sim 150\mev$ where collisions dominate the evolution of the
neutrino system and matter and thermal effects become relevant. As a result, the sterile neutrino relic density depends
quadratically on $m_s$ and $\kev$ sterile neutrinos with $\sstt\sim
10^{-8}$ are required to account for the dark matter. In models with low
reheating temperature, on the other hand,  $\Omega_{\nu_s}$ depends linearly
on $m_s$ and much larger mixing angles are required to explain
the dark matter.

Cosmological and
astrophysical observations can be used to constrain sterile neutrinos as  dark
matter candidates. The observed clustering  on small scales of the
Lyman-$\alpha$ forest, for instance,
constrains the sterile neutrino mass from below. To obtain a limit on $m_s$,
the flux power spectrum of the Lyman-$\alpha$ forest must be carefully
modeled using numerical simulations. The analysis presented in
\cite{Seljak:2006qw} and \cite{Viel:2006kd}
respectively cite $m_s>10\kev$ and $m_s>14\kev$ as their limits, though a $30\%$
discrepancy between them still exists. Such bounds, however, were obtained for
sterile neutrinos produced in the standad cosmological model and do not direcly apply to
the scenario we consider. That is why  we will be mainly
concerned with another bound, that derived from X-rays measurements. Sterile neutrinos may
radiatively decay through $\nu_s\to \nu_\alpha + \gamma$ producing a
monoenergetic photon, $E_\gamma =m_s/2$. X-ray observations may therefore be
used to 
constrain or infer the mass of the sterile neutrino. In a recent analysis of
the X-ray background from HEAO-1 and XMM-Newton, for example, the
following limit 
\begin{equation}\label{eq:lim}
\sstt< 1.15\times 10^{-4}\left(\frac{m_s}{\kev}\right)^{-5}\left(\frac{0.26}{\Omega_{\nu_s}}\right)
\end{equation}     
relating $\sstt$, $m_s$ and $\Omega_{\nu_s}$ was found
\cite{Boyarsky:2005us}. This bound is model independent, it applies to both
the standard production mechanism and to the production in models with low
reheating temperatures.

In figure \ref{fig:10b} we display the sterile neutrino relic density as  a
function of $\sstt$ for different values of $\tr$  and $m_s=1\kev$. The limit
from X-rays, equation (\ref{eq:lim}), is also shown and rules out the
upper-right part of the figure. The different lines represent different
reheating temperatures. Notice, for instance, that $\tr=4\mev$,
$\Omega_{\nu_s}=\Omega_{dm}$ is not a viable point of the parameter space as
it is incompatible with the X-rays limit. Indeed,  sterile neutrinos  can
account for the dark matter only if $\tr\gtrsim 7\mev$.

Turning this argument around we can also say that if dark matter consists of
sterile neutrinos, they provide the strongest constraint on the reheating
temperature. The present bound,
in fact, gives $\tr\gtrsim 2-4\mev$ and is based on the effect of active neutrinos on
big bang nucleosynthesis. Dark matter sterile neutrinos might yield a more
stringent constraint. Finally, notice that this bound on $\tr$ was obtained for
a sterile neutrino with $m_s=1\kev$ but it only becomes stronger for larger
 masses. Dark matter sterile neutrinos, therefore, are useful probes of the
 early Universe.

\section{Sterile neutrino production in oscillations and decays}\label{sec:5}

The  field $\phi$ responsible for the reheating process may also
 have a direct decay mode into sterile neutrinos ($\phi\to \nu_s\nu_s$), opening an additional
 production mechanism for $\nu_s$. As we will see, this mechanism significantly alters the predictions
 obtained in the previous section. In \cite{Shaposhnikov:2006xi}, the
 production of sterile neutrinos in inflaton decays was investigated, but
 not in the context of low reheating temperatures. The main motivation to
 consider this mechanism is the
 conflict between the constraints from X-ray observations and those from
 small-scale structure that rule out the minimal production scenario for
 sterile neutrino dark matter. 

\EPSFIGURE[t]{spevol,width=10cm}{The evolution of the sterile neutrino energy
  spectrum for $\tr=4\mev$, $b=10^{-3}$ and $\sstt=10^{-8}$.\label{fig:10}}

As mentioned in section 2, the decay $\phi\to \nu_s\nu_s$ gives the following
contribution to the sterile neutrino collision integral
\begin{equation}
C_{\phi\to \nu_s\nu_s}= b\frac{2 \pi^2}{(m_\phi/2)^2}\Gamma_\phi n_\phi
\delta(p-m_\phi/2)\,,
\label{eq:stcol}
\end{equation}
where $b$ denotes the $\phi$ branching ratio into sterile neutrinos, and $m_\phi$,
$n_\phi$ are respectively the $\phi$ mass and number density. Being $\phi$
non-relativistic, each $\nu_s$ is born with momentum $p=m_\phi/2$, as enforced
by the delta function. Due to this new contribution, $f_{\nu_s}$ will  now depend not only on $\tr, m_s,$ and $\sstt$ but
also on $b$ and $m_\phi$. To keep things simple we will set $m_\phi=100\mev$
and  study the dependence of the different observables with $b$.

\DOUBLEFIGURE[t]{omesintr4,width=7.5cm}{omesintr4B,width=7.5cm}{The sterile neutrino relic density as a
  function of $\sstt$ for $\tr=4\mev$. The sterile neutrino mass is set to
  $1\kev$ and the curves correspond to two different values of $b$. The bound
  from X-rays observations is also shown.\label{fig:11}}{The sterile neutrino relic density as a
  function of $\sstt$ for $\tr=4\mev$. The sterile neutrino mass is set to
  $10\kev$ and the curves correspond to two different values of $b$. The bound
  from X-rays observations is also shown.\label{fig:12}}

Figure \ref{fig:10} displays the evolution of the sterile neutrino energy spectrum for $\tr= 4\mev$, $b=10^{-3}$, and $\sstt=10^{-8}$. Each line
corresponds to a different temperature. It is not
difficult to decipher what is going on. Whenever a $\phi$ decays, a peak at
$p=m_\phi/2$ in  $f_{\nu_s}$ is generated. But not all $\phi$'s decay at the
same time. And the momentum of the sterile neutrinos produced in earlier
decays is redshifted when later decays occur. That is why, at any given
temperature, the resulting spectrum has a drastic jump at $p\sim m_\phi/2$, with  all
the neutrinos produced before (in decays) lying at  smaller momenta. As we
approach the radiation dominated epoch, the redshift essentially ceases and
only residual decays modify  the spectrum at large $p/\tg$. At the end, no traces of the
discontinuity at $p=m_\phi/2$ are left in the primordial spectrum.

The sterile neutrino relic density is shown in figure \ref{fig:11} as a
function of $\sstt$. For that figure $\tr=4\mev$, $m_s=1\kev$ and the
two curves correspond to $b=10^{-2}$ and $b= 10^{-3}$. The solid line is the X-ray constraint
 obtained from equation (\ref{eq:lim}). The relic density
behaves in a similar way for the different values of $b$.  At large mixing
angles, the production of sterile neutrinos is dominated by oscillations and
independent of $b$. That is the case we dealt with in the previous section. At smaller mixing angles,  we encounter
an intermediate region where both production mechanisms are relevant and the
relic density depends on $b$ and $\sstt$. Finally, at even smaller mixing
angles, sterile neutrinos are produced dominantly in $\phi$ decays and
therefore the relic density does not depend  on $\sstt$, as signaled by the
horizontal lines observed in figure \ref{fig:11}. In that region the sterile
neutrino relic density is simply proportional to $b$. If sterile neutrinos
account for the dark matter, $\Omega_{\nu_s}=\Omega_{dm}$, the X-rays constraint
requires a small mixing angle, $\sstt\lesssim 10^{-4}$.

New viable regions, where the sterile neutrino is produced in $\phi$ decays
and makes up
the dark matter of the Universe, can be read off figures \ref{fig:11} and \ref{fig:12}. For instance, a $m_s=1\kev$ sterile neutrino
with $\sstt< 10^{-4}$ will be a good dark matter
candidate for $\tr \sim 4\mev$ and $10^{-3}<b<10^{-2}$. For decay-dominated
production, $\Omega_{\nu_s}$ is simply proportional to $\tr$, 
\begin{equation}
\Omega_{\nu_s}\propto bm_s\tr\,. 
\label{eq:scal}
\end{equation}
Using this equation in conjuntion with figures \ref{fig:11} and \ref{fig:12},
additional allowed regions can be found.

Figure \ref{fig:12} is analogous to figure \ref{fig:11} but for a larger value of
the sterile neutrino mass, $m_s=10\kev$. The two curves correspond to
$b=10^{-3}$ and $b=10^{-4}$. Owing to the increase in $m_s$, the X-ray limit
becomes much stronger than in figure \ref{fig:11}. Indeed, it constrains
dark matter sterile neutrinos to have a very small mixing angle,
$\sstt\lesssim 10^{-9}$. 

In the standard production mechanism, such small mixing angles are not allowed
as they yield a too small  sterile neutrino relic density,
$\Omega_{\nu_s}\propto \sstt$.  For  sterile neutrinos originating in $\phi$
decays, on the contrary,
the production mechanism and  the radiative
decay  are controlled by two different parameters. In fact,
$\Omega_{\nu_s}\propto b$ whereas $\Gamma(\nu_s\to
\nu_\alpha+\gamma)\propto \sstt$. Thus, no
matter how small  $\sstt$ -and consequently $\Gamma(\nu_s\to
\nu_\alpha+\gamma)$- is, it is still possible to find appropriate values of
$b$, $\tr$ and $m_s$  such that $\Omega_{\nu_s}=\Omega_{dm}$. In other words,
for $b\neq 0$ the X-rays limit can always be satisfied.

\section{Conclusions}
We numerically studied the production of sterile neutrinos in models with low reheating
temperatures. Two  production mechanisms for the sterile neutrinos were taken
into account:  active-sterile neutrino oscillations
($\nu_\alpha\leftrightarrow \nu_s$) and the direct decay of the scalar field
($\phi\to \nu_s\nu_s$). Several observables, including $f_{\nu_s}$, $N_\nu$,
 and $\Omega_{\nu_s}$, were computed  for different sets of reheating and mixing
parameters. We showed that in these models, LSND  sterile neutrinos may still give
$N_\nu\sim 3$ --avoiding problems with big bang nucleosynthesis-- and that $\kev$
sterile neutrinos may  account for the dark
matter of the Universe. Dark matter sterile neutrinos produced in oscillations
were found to be effective probes of the early Universe, as they constrain the reheating
temperature to be rather large, $\tr\gtrsim 7\mev$. Finally, we showed that
sterile neutrinos  originating in decays may explain the dark matter and
satisfy the bounds from X-ray observations.   

\acknowledgments
I would like to thank A. Kusenko and G. Gelmini for their comments and useful suggestions.

\end{document}